\documentclass[aps,prb,twocolumn,groupedaddress,showpacs]{revtex4}
\usepackage{graphicx}
\newcommand{\bkp}{\mbox{\boldmath $k_{\parallel}$}}

\newcommand{\bsig}{\mbox{\boldmath $\sigma$}}
\newcommand{\bj}{\mbox{\boldmath $j$}}

\begin{document}
\title{THEORY OF SPIN-TRANSFER TORQUE IN THE CURRENT-IN-PLANE GEOMETRIES}

\author{O. Wessely}
\affiliation{Department of Mathematics, Imperial College, London SW7 2BZ, United
 Kingdom}
\affiliation{Department of Mathematics, City University,London EC1V 0HB, United 
Kingdom}
\author{A. Umerski}
\affiliation{Dept of Mathematics, Open University, Milton Keynes MK7 6AA, United Kingdom}
\author{J. Mathon}
\affiliation{Department of Mathematics, City University,London EC1V 0HB, United 
Kingdom}

\date{\today}

\begin{abstract}
Two alternative current-induced switching geometries, in which the current flows parallel to the magnet/nonmagnet 
interface, are investigated theoretically using the nonequilibrium Keldysh theory. In the first geometry, 
the current is perpendicular to 
the polarizing magnet/nonmagnet interface but parallel to the nonmagnet/switching magnet interface (CPIP). 
In the second
geometry, the current is parallel to both the polarizing magnet/nonmagnet and nonmagnet/switching magnet
interfaces (CIP). Calculations for a single-orbital tight binding model indicate that the spin current  
flowing parallel to the switching magnet/nonmagnet interface can be absorbed by a lateral switching magnet 
as efficiently as in the traditional current-perpendicular-to-plane (CPP) geometry. The results of the model 
calculations are
shown to be valid also for
experimentally relevant Co/Cu CPIP system described by fully realistic tight binding bands fitted to 
an ab initio band structure. It is shown that almost complete absorption of the incident spin current 
by a lateral switching magnet occurs when the lateral dimensions of the switching magnet are
of the order of 50-100 interatomic distances, i.e., about 20nm and its height as
small as a few atomic planes. It is also demonstrated
that strong spin current absorption in the CPIP/CIP geometry is not spoilt by the presence of a rough interface
between the switching magnet and  nonmagnetic spacer. Polarization achieved using a lateral magnet in the CIP
geometry is found to be about 25\% of that in the traditional CPP geometry. The present CPIP calculations of the spin
transfer torque are also relevant to the so called
pure-spin-current-induced magnetization switching that had been recently observed.
\end{abstract}


\maketitle
\section{Introduction}
\noindent In experiments on current induced switching of magnetization (see e.g. Ref.~\onlinecite{ref1}),
current passing through a thick polarizing magnet (PM) becomes
spin polarized. The spin polarized current (spin current) then flows through a
nonmagnetic layer (the spacer layer) and becomes partially or fully absorbed by
a switching magnet (SM). The absorbed spin current
exerts a spin-transfer torque on the switching magnet and this torque can be
used to  switch the direction of the magnetization of the
switching magnet between the parallel (P) and antiparallel (AP) orientations
relative to the magnetization of the polarizing magnet. In this traditional
setup  the current is perpendicular to both the PM/spacer and spacer/SM
interfaces. This setup is referred to as current perpendicular to plane (CPP) 
geometry and is shown schematically in Fig.1.
\begin{figure}
\includegraphics[width=0.35\textwidth]{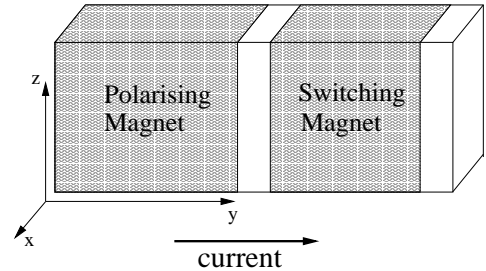}
\caption{\footnotesize CPP switching geometry.} 
\end{figure}
The switching process 
relies on the scenario in which one of the configurations (P or AP) becomes
unstable, at a critical current, the other configuration is stable and,
therefore, available for switching into. However, in the presence of an
external magnetic field stronger than the coercive field of the switching
magnet, it is found experimentally \cite{ref2,ref3,ref4,ref5} that, for current greater than a critical
value and with the correct sense, neither the P nor the AP configuration is
stable. The magnetization of the switching magnet then precesses continually 
and becomes a source of microwave generation. It was also proposed
\cite{ref51} 
that microwave generation can occur even in the absence of an applied field
provided the spin-transfer torque has both the in-plane and out-of plane components
of appropriate relative sign.
Both the switching and microwave generation scenarios have potentially very important
applications. However, to limit the current to acceptable values and to minimize the
Oersted fields generated by the current, experiments are performed on CPP nanopillars
with a very small diameter of the order of 100nm. Such nanopillars are difficult to 
prepare. Moreover, to achieve a usable microwave power, large arrays of CPP 
nonopillars would have 
to be manufactured, and this is even more difficult to achieve. We have, therefore,
investigated theoretically two alternative geometries, shown in Fig.2, which may have
\begin{figure}
\includegraphics[width=0.5\textwidth]{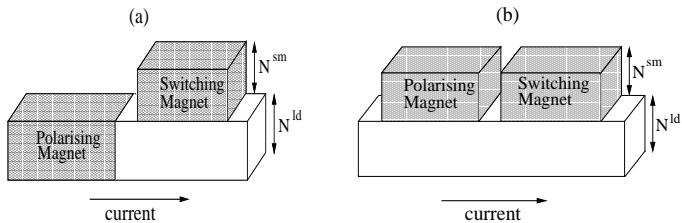}
\caption{\footnotesize CPIP (a) and CIP (b) switching geometries.} 
\end{figure}
interesting applications since they offer
much more flexibility for design of current-induced switching and microwave 
generation devices. 

In the first geometry shown in Fig.2a, the current is perpendicular to 
the PM/spacer interface but parallel to the spacer/SM interface (CPIP). In the second
geometry shown in Fig.2b the current is parallel to both the PM/spacer and spacer/SM
interfaces (CIP). It is clear from Fig.2 that switching magnets in the CPIP and CIP geometries
are
arrays of either magnetic dots or wires deposited on the surface of a nonmagnetic substrate.
It should be noted that our CPIP geometry in which the current flows parallel to the 
switching magnet/nonmagnet
interface is closely related to that used in the so called pure-spin-current-induced 
magnetization switching  which was recently demonstrated experimentally \cite{ref55}.
This is because, just like in the pure spin current switching, no net charge current flows in the CPIP
and CIP geometries through the switching magnet in the direction perpendicular 
to its interface with the spacer. Nevertheless we shall see that a spin current is absorbed by
the switching magnet and this gives rise to a nonzero spin-transfer torque. This effect 
is sometimes called nonlocal spin-transfer torque (for detailed discussion of spintronics circuits see 
Ref.~\onlinecite{bauer}).
 
While the potential advantages of the CPIP and CIP geometries are obvious the crucial question 
is whether these alternative geometries are as efficient for switching/microwave generation as
the traditional CPP geometry. To address this question we have applied the nonequilibrium
Keldysh formalism \cite{ref6,ref7,ref8} to calculate from first principles 
the spin-transfer
torques in the CPIP and CIP geometries. We assume in all our calculations that the spin diffusion
length is much longer than the dimensions of our system (spin is conserved).
We performed  calculations of the spin-transfer torque for perfect CPIP and CIP systems
(ballistic limit) and also in the case of a rough nonmagnet/magnet interface to check that our 
results remain
valid beyond the ballistic limit.  Rather surprisingly both our single-orbital model 
calculations and fully realistic
calculations for Co/Cu show that the spin current flowing parallel to the spacer/SM interface 
can be absorbed by the switching magnet as efficiently as in the traditional CPP geometry. 
Spin polarization of the current in the CIP geometry is not as large as in the CPP geometry
but remains sizable, of the same order of magnitude as in the CPP geometry. 
%
\section{Theoretical formulation}
%
\noindent
The Keldysh formalism had been applied previously by Edwards {\it et al.} \cite{ref8} 
to calculate the spin-transfer torque in the CPP geometry. An essential requirement for the
implementation of the Keldysh formalism is that a sample with an applied bias can be cleaved 
into two noninteracting
left (L) and right (R) parts by passing a cleavage plane between two neighboring atomic planes. 
It follows that, initially, neither charge nor spin current flows in the cleaved system although the
left and right parts of the sample have different chemical potentials. 
This is most easily achieved for a tight-binding (T.-B.) band structure since the T.-B. hopping 
matrix between the L
and R parts can be switched off. We shall, therefore, describe our systems by a tight-binding model, in
general multiorbital with s,p, and d orbitals whose one-electron parameters are fitted to 
first-principle band structure, as described previously \cite{ref9}. 
The hopping between the L and R parts is then turned on 
adiabatically and the
system evolves to a steady state. The nonequilibrium Keldysh formalism provides a
prescription for calculating the steady-state charge and spin currents flowing between the L 
and R parts of the connected sample in terms of local one-electron Green functions
for the equilibrium cleaved system. In the CPP geometry, considered by Edwards {\it et al.} \cite{ref8}, 
the sample 
is translationally invariant in the direction parallel to all the interfaces and, therefore,  
the relevant quantity is the total spin current flowing between any two 
neighboring atomic planes. In particular, the spin-transfer torque acting on the switching magnet 
is obtained as the difference
between the spin currents entering and leaving the switching magnet (the spin current is naturally
conserved in the nonmagnetic spacer and leads). Edwards {\it et al.} \cite{ref8} showed that the local spin
current is expressed entirely in terms of the  
one-electron surface Green functions $g_{L}(\bkp)$ and $g_{R}(\bkp)$ for the cleaved sample. Here,
$\bkp$ is the wave vector parallel to the interface. 
The  Green
functions at the surfaces of the cleaved system are obtained from the surface Green functions 
of the nonmagnetic leads by the method of adlayers \cite{ref9}. In this method one "grows" 
the sample by depositing, one by one,
all its atomic planes on the leads and, after each deposition, the surface Green
function is updated using Dyson's equation. The surface Green function of semi-infinite leads is obtained
by the method of Umerski \cite{andrey}.

We now wish to apply the Keldysh method to the CPIP and CIP geometries. Referring to Fig.2, it is clear
that the translational invariance is broken in the z and y directions but $k$-space description remains
valid in the x direction. We, therefore, need to work in a representation that is atomic-like in 
the z and y directions but Bloch-like in the x direction. The method for modelling CPIP and CIP systems 
is shown schematically in
Fig.3 for the CPIP geometry. The whole system is built up from chains of atoms parallel to the z axis
\begin{figure}
\includegraphics[width=0.45\textwidth]{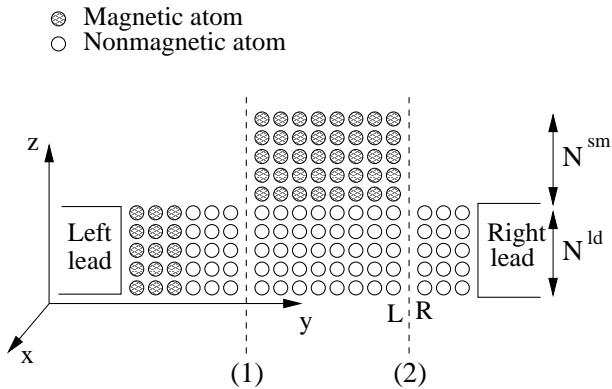}
\caption{\footnotesize Schematic model of the CPIP geometry. Two alternative locations of a cleavage plane are 
labeled by (1) and (2).} 
\end{figure}
which are repeated periodically in the x direction. We shall label the position of each chain by 
$n$ and the position of atoms within a chain by $m$. Although we shall frequently refer to chains, 
in reality each chain stands for a sheet of atoms since the chains are repeated periodically in the x
direction.
The tight-binding on-site potentials depend on the location of each atom in the sample and those
for magnetic atoms  include an interaction between
electrons in d orbitals which leads to an exchange splitting of the bands in the ferromagnets.
The region which lies outside the sample is modelled by fictitious 
atoms with an infinite on-site potential which prevents electrons from hopping to these vacant sites.
All chains can be thus regarded as having the same length of $N=N^{ld}+N^{sm(vac)}$ atoms, where 
$N^{ld}$ 
and $N^{sm(vac)}$ are, respectively, the numbers of atoms in the lead and in the switching magnet 
(vacuum) in the 
vertical z direction. It follows that we can create the whole sample by depositing all its chains
one by one on semi-infinite left and right leads. 
The surface Green functions on the chains located immediately to the
left and right of a cleavage plane, that are required in the calculation of the spin current, 
are obtained by updating the Green function  
from the Dyson equation after each chain deposition. Since the deposition of chains of atoms takes place 
in real space in the z direction the Green function is a matrix of dimension
$(2\times N^{orb}\times N)\times (2\times N^{orb}\times N)$, where $N$ is the number of atoms in a chain and 
$N^{orb}$ is the number of orbitals. The factor 2 appears because the Green function has two
components corresponding to two spin projections on the spin quantization axis.

To calculate the spin and charge currents we
assume that a bias $V_{b}$ is applied between the left and right leads. 
Our goal is to determine the spin and charge currents between any two neighboring chains of atoms 
parallel to the the z axis, i.e. to the interface between the
left (polarizing) magnet and the lead. If the cleavage line is first passed to the
left of the switching magnet and then to the right of the magnet, as indicated in Fig.3, the 
spin-transfer torque acting on the switching magnet is obtained as the difference between 
the total spin 
currents in these two locations. Following Edwards {\it et al.} \cite{ref8} and assuming the 
linear-response case of a small bias, it is straightforward 
to show that the thermal average of the total spin current $j_{n-1}$ flowing between the 
chains $n-1$ and $n$ is given by
\begin{widetext}
\begin{eqnarray}
<\bj _{n-1}> = \frac{1}{4\pi}\sum_{k m}  \, Re \, Tr \{[g_{L}TABg^{\dagger}_{R}
T^{\dagger}-AB+\frac{1}{2}(A+B)]\bsig \}_{mm}V_{b},
\label{eq1}
\end{eqnarray}
\end{widetext}
where $A=[1-g_{L}^{\dagger}Tg_{R}^{\dagger}T^{\dagger}]^{-1}$, 
$B=[1-g_{L}Tg_{R}T^{\dagger}]^{-1}$ are defined in terms of retarded surface Green function
matrices
$(g_{L})_{mm'k}$, $(g_{R})_{mm'k}$ for the decoupled equilibrium system. The subscript L(R) refers
to the chains on the left (right) of the cleavage line. The Green functions depend on the wave
vector $k$ labelling  Bloch states in the x direction and on the indices $m$, $m'$ labelling the atoms
in a chain. The matrix $T$ is the tight-binding interchain hopping matrix. The components of 
$\bsig$ are direct products of the 
2$\times$2 Pauli matrices $\sigma_{x}$, $\sigma_{y}$, $\sigma_{z}$ and $(N\times N^{orb})\times (N\times N^{orb})$ 
unit matrix. 
Finally, the trace in Eq.(1) is taken over all the orbital and spin indices which are suppressed.
Equation (1) yields the charge current if $\frac{1}{2}\bsig$ is replaced by a unit matrix multiplied
by $e/\hbar$, where $e$ is the electronic charge.

It follows from Eq.(1) that the total spin current (charge current) between the chains $n-1,n$ is the
sum of partial currents flowing between pairs of atoms which are located on the opposite sides 
of the cleavage plane and connected by the T.-B. hopping matrix. By evaluating the individual partial
currents we can, therefore, obtain a detailed information about the local current flow. Equation (1)
yields, of course, only information about current flow in the y direction, which is perpendicular 
to the cleavage line. However, by applying locally Kirchhoff's law, the current components in the
direction parallel to the cleavage line (z axis) can  also be determined. The current vector
describing the flow of charge current between any two neighboring atoms in the (y,z) plane can be
thus reconstructed. While local currents are not conserved, the total charge current between any
two neighboring chains anywhere in the system is, of course, conserved. The total spin current between
neighboring chains is conserved in the nonmagnetic parts of the system but can be absorbed in the
magnets, which gives rise to spin-transfer torque. The application of Eq.(1) to specific CPIP and CIP
structures will be discussed in Section 3. 
%
\section{Results for a single-orbital tight binding model.}
%
\noindent
To gain some insight, we have first applied the Keldysh formalism to the CPIP and CIP geometries using 
a single-orbital tight-binding model with atoms on a simple cubic lattice and nearest-neighbour hopping $t$. 
In this model the relevant parameters are the on-site potentials $V^{\uparrow}$, $V^{\downarrow}$ which 
are measured in the units of $2t=1$. The Fermi level is always set at zero.

We begin with the CPIP geometry illustrated in
Fig.2a. For a meaningful comparison of the CPIP geometry with the traditional CPP setup, we also need to 
determine the CPP spin current for a system which is finite in the z direction. We, have therefore,
applied to the CPP geometry the same real space method described in section 2 for the CPIP and 
CIP geometries. 
We choose the total number $N$ of atoms in a chain to be the same in the CPP and CPIP geometries 
and make all the spin currents dimensionless by dividing them by the 
total charge current multiplied by $\hbar/2e$, where $e$ is the electronic charge. The magnetization of the polarizing magnet is assumed to be parallel to the x axis and
that of the switching magnet is parallel to the z axis. For simplicity, we choose the polarizing magnet 
to be semi-infinite in the y
direction. The switching magnet should, of course, be finite since the torque is calculated by taking the
difference between the spin currents before and after the switching magnet. However, it has been
demonstrated for the CPP geometry \cite{ref51} that the dependence of the outgoing spin current on the
switching magnet thickness is almost exactly the same as the dependence of the spin current 
on the distance 
from the spacer/switching magnet interface in a semi-infinite
magnet. We checked that this is also true for the CPIP geometry. We may, therefore, determine the
spin-transfer torque using a semi-infinite switching magnet. The advantage of using a semi-infinite magnet
is a faster convergence of the $k$-space sum since small and physically unimportant interference effects 
which occur in a ferromagnet of a finite thickness are eliminated. 
 
Placing a cleavage plane in the position (1) in Fig.3, we first determine from 
Eq.(1) the spin 
current in the nonmagnetic spacer, i.e. the spin current incident on the switching magnet. 
We then place a cleavage
plane between any two neighboring atomic chains in the switching magnet and determine again from Eq.(1)
the local spin current in the magnet. The spin current $j_{n-1}$ flowing between the 
chains $n-1$ and $n$ can be then plotted as a function of the position $n$ of the cleavage plane in the
switching magnet. Such plots are shown in Fig.4 for $N=20$ and for three different aspect ratios
$N^{sm}/N=1/20$, $N^{sm}/N=10/20$, and $N^{sm}/N=19/20$ corresponding to the height of the switching magnet in the
CPIP geometry of one atom, ten atoms, and nineteen atoms. The dependence of the spin current on $n$ in the
CPP geometry is also shown in Fig.4. The spin current curves in Fig.4a and 4b correspond to different 
tight-binding on-site potentials in the polarizing and switching magnets, which are listed in the figure. 
Those in Fig.4a were chosen so that 
the Fermi level in the polarizing and switching magnets intersects both the majority- and minority-spin bands
(a weak magnet)
and there is a perfect matching between the bands of the nonmagnetic spacer and one of the ferromagnet bands. 
In Fig.4b both the polarizing and switching magnets are half-metals, i.e., the minority-spin band is empty.
%
%
It should be noted that, in general, the spin current relevant for
current-induced switching has an in-plane (x) and out-of-plane (y) components \cite{ref8}. However, we 
show in Fig.4 only the in-plane component since it is usually most important in switching.
\begin{figure}
\includegraphics[width=0.45\textwidth]{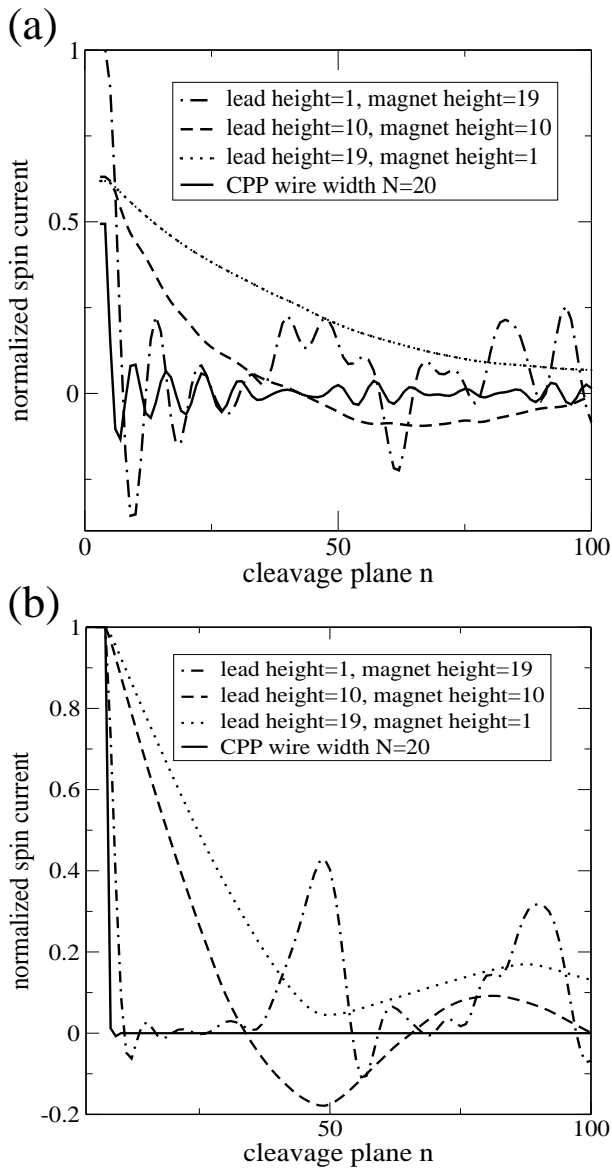}
\caption{\footnotesize Dependence of the spin current on the position $n$ of the cleavage plane in the
switching magnet. The on-site potential parameters in (a) are
$V^{\uparrow}=1.5$, $V^{\downarrow}=2.5$ for the PM, $V^{\uparrow}=V^{\downarrow}=1.7$ for the spacer, and
$V^{\uparrow}=1.7$, $V^{\downarrow}=2.4$ for the SM. 
The on-site potential parameters in (b) are
$V^{\uparrow}=0.7$, $V^{\downarrow}=4$ for the PM, $V^{\uparrow}=V^{\downarrow}=0.7$ for the spacer, and
$V^{\uparrow}=0.7$, $V^{\downarrow}=4$ for the SM.} 
\end{figure}
It can be seen from Fig.4 that both the CPP and CPIP spin currents decrease as the cleavage plane
is moved
through the switching magnet and become almost zero for a switching magnet of about fifty to hundred
chains wide. The only exception occurs for the aspect ratio $N^{sm}/N=19/20$  for which the
spin current is virtually nondecaying. This will be explained later, once the physical mechanism governing the 
spin current absorption is clarified. 

Zero outgoing spin current corresponds to complete absorption of the spin current by the
switching magnet, i.e., maximum spin-transfer torque. Fig.4 demonstrates that almost complete absorption 
of the spin current is achieved not
only in the CPP but also in the CPIP geometry. 

It should be noted that the rate of decay of the CPIP spin current for a half-metallic magnet (Fig.4b) is 
comparable to that
for a weak magnet (Fig.4a) but the CPP spin current decays much faster in a half-metallic ferromagnet. 
 
Since the results in Figs.4a and 4b were obtained for
magnets with different band parameters, it is clear that a complete absorption of the spin current by the
switching magnet in the CPIP geometry is a general phenomenon. It can  also be seen from Fig.4 that a
switching magnet of height of only one atom has essentially the same absorbing power as that having 
height of ten atoms. 

To understand these rather surprising results, we first
recall the physical mechanism that governs the absorption of spin current in the CPP geometry \cite{stiles,ref51}. For noncolinear magnetizations of the polarizing and switching magnets, the spin of electrons 
incident on the switching magnet is
at an angle to its exchange field. It follows that the spin 
must precess in the exchange field of the switching magnet. The precession
frequency is determined by the components of the wave vectors of majority- and minority-spin 
electrons parallel to the current flow (perpendicular to the interfaces). 
Given that the sum of the energies corresponding to perpendicular and parallel motion of electrons is constant 
(equal to the Fermi energy), the perpendicular components of the wave vector, which determine the precession
frequency, are functions of the parallel component $\bkp$.
Since the total spin current involves the sum over $\bkp$, destructive interference of precessions with
different frequencies occurs. The conventional stationary phase argument \cite{stiles} then shows  that 
only an extremal frequency of spin current oscillations survives. The stationary phase argument also
predicts that the amplitude of spin current oscillations decays as a function of 
the distance from the spacer/magnet interface. Such a behaviour of the CPP spin current is clearly 
seen in Fig.4a. The fast decay of the CPP spin current in the case of a half-metallic switching magnet can
be explained as follows. The wave function of an electron with a spin at an angle to the exchange field of a
half-metallic switching magnet is a linear combination of the wave functions with spin parallel and
antiparallel to the exchange field. However, since only electrons with one spin projection on the direction of
the exchange field (magnetization) exist in a half-metallic magnet the precession amplitude must decay
exponentially. This is the behaviour seen for the CPP spin current in Fig.4b. 
  
It is reasonable to assume that spin precession mechanism is also responsible for the decay of the spin
current in the CPIP geometry. However, we need to establish that destructive interference of precessing spins
can occur in this geometry and also that electrons travelling parallel to the spacer/switching
magnet interface do penetrate the switching magnet, so that their spin can precess in the local 
exchange field. In an inhomogeneous finite sample shown in Fig.3, size quantization occurs and
electrons thus travel in discrete size-quantized conductance channels. This effect
combined with the sum over the wave vector $k$ in the x direction provides in the CPIP geometry 
the relevant channels for destructive interference. However, because of the complexity of size
quantization both in the y and z directions, a simple stationary phase argument is no longer applicable
and an analytical formula for the spin current decay in the CPIP geometry is thus not available.
The only exception is the case with an aspect ratio $N^{sm}/N=19/20$ in Fig.4a where size quantization is so
severe that only one conductance channel is available. Destructive interference then occurs due only to 
different $k$-space channels to which the conventional stationary phase argument is applicable. In contrast to
the planar CPP geometry, the $k$-space sum in the CPIP geometry is onedimensional and, therefore, the decay
of spin current oscillations is much slower then in the planar CPP geometry. 

Although in the general case of a large number of size-quantized conductance channels we do not have a simple
stationary-phase formula for the spin current in the CPIP geometry, we can nevertheless make an estimate 
of the slowest decay of the spin current in a lateral switching magnet. The spin current in Eq.(1) is the trace
over the real-space position in the vertical (z) direction combined with the sum over the wave vector $k$
labelling Bloch states in the x direction. The
trace in the real space is essentially equivalent to a sum over discrete size-quantized conductance channels.
For each conductance channel the sum over the wave vector $k$ can be performed using the conventional
stationary phase argument (see Ref.~\onlinecite{itoh}). That gives a decay of the spin current in each discrete 
conductance channel  
of the form $\propto 1/\sqrt{n}$, where n is the position of the cleavage plane in the switching magnet.
Since this conventional stationary-phase argument can be applied to each conductance channel, the slowest decay
of the spin current must be $\propto 1/\sqrt{n}$. In practise, destructive interference between different conductance
channels also occurs, and that should lead to a faster decay than the most pessimistic estimate
$\propto 1/\sqrt{n}$. 
 
It remains to demonstrate that transport electrons penetrate the switching magnet
despite the fact that they travel parallel to the interface. To show that we have determined the distribution 
of the local charge current in the switching magnet using the method outlined in section 2. The behaviour of the
charge current is shown in Fig.5 for $k=0$ (strictly two-dimensional system) and the aspect ratio  $N^{sm}/N=10/20$.
\begin{figure}
\includegraphics[width=0.5\textwidth]{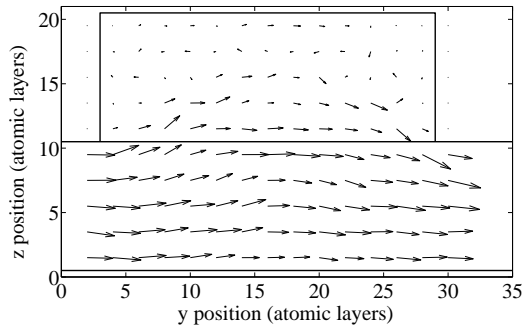}
\caption{\footnotesize Distribution of the charge current in the
switching magnet in the CPIP geometry. The on-site potential parameters are
$V^{\uparrow}=0.7$, $V^{\downarrow}=3.7$ for the PM, $V^{\uparrow}=V^{\downarrow}=1.1$ for the spacer, and
$V^{\uparrow}=1.1$, $V^{\downarrow}=1.9$ for the SM.} 
\end{figure}
The orientation of each arrow in Fig.5 represents the direction of the current flow and the length of the arrow
gives the magnitude of the local charge current flowing between neighboring atoms. Figure 5 demonstrates 
that there is strong penetration of transport electrons into the switching
magnet, and it is the spin precession of these electrons that results in a spin-transfer torque 
(spin current absorption) which is as large as in the CPP geometry. 

Finally, we need to explain why the decay of the CPIP spin current in a half-metallic ferromagnet is slower 
than in the CPP geometry. In the CPP geometry all electrons have to pass through the switching magnet and the
spin current thus decays exponentially as discussed above. In the CPIP geometry there are many electrons that
penetrate only partially the switching magnet and are then reflected back to the spacer. The spin of 
such electrons with a shallow penetration can precess in the exchange field of the switching magnet 
and the decay of the spin current is
thus not qualitatively different from that for a weak magnet (see Fig.4a and 4b). 

The results shown in Fig.4 and Fig.5 are for structures with perfect interfaces, that are illustrated in Fig.2a.
Interfaces in real structures may well be rough and it is, therefore, necessary to investigate the effect of
interfacial roughness on the absorption of the spin current by the lateral switching magnet. Since the systems
we consider are 
"grown" in real space it is straightforward to include in our calculations the effect of a random intermixing 
of atoms in the nonmagnetic spacer and switching magnet. The effect of an intermixing over two interfacial atomic
planes on the absorption of the spin current is shown in Fig.6. The intermixing was modelled by replacing the two
interfacial atomic planes by a 50\% alloy of spacer and magnet atoms. The results for a perfect system 
are also reproduced in Fig.6. It can be seen that intermixing does not spoil the strong absorption of the spin current 
by a lateral switching magnet. The other interesting feature is that the spin current for a perfect CPIP system
exhibits oscillations reminiscent of those that are seen in the CPP geometry. While oscillations of the spin
current in the CPP geometry can be explained by the stationary phase theory, a simple stationary-phase argument is 
not available for the CPIP geometry and the precise origin of the oscillations in this geometry is thus not clear.
However, it can be seen in Fig.6 that CPIP oscillations are removed in a system with rough interface.
\begin{figure}
\includegraphics[width=0.45\textwidth]{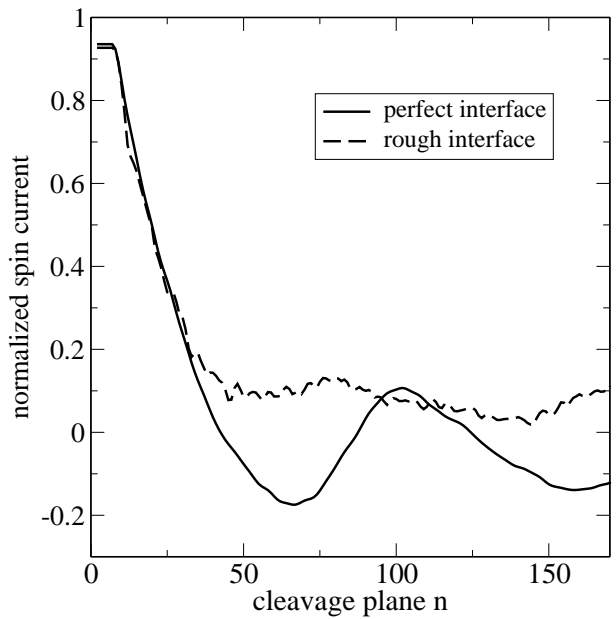}
\caption{\footnotesize Dependence of the spin current on the position $n$ of the cleavage plane in the
switching magnet for a rough and a perfect interface. The on-site potential parameters are $V^{\uparrow}=2.1$, 
$V^{\downarrow}=2.9$
for the PM ,$V^{\uparrow}=V^{\downarrow}=2.1$ for the spacer, and
$V^{\uparrow}=2.1$, $V^{\downarrow}=2.9$ for the SM. 
The lead/magnet height is 10/10 atomic planes for both the perfect and the rough systems} 
\end{figure}

We now investigate the CIP geometry in which the current flows parallel not only to the interface between the
spacer and the switching magnet but also to the interface between the
spacer and the polarizing magnet. Since the absorbing power of the switching magnet in the CIP geometry
must clearly be the same as in the CPIP geometry, the key question here is the polarizing ability of 
a polarizing magnet whose interface with the spacer is parallel to the current flow. To determine the spin current, we
proceed as in the CPIP geometry (Fig.3). We place a cleavage
plane between any two neighboring atomic chains in the switching magnet and determine from Eq.(1)
the local spin current as a function of the position $n$ of the cleavage plane in the
switching magnet. The continuity of the spin current guarantees that the value of the spin 
current at the spacer/switching magnet interface is equal to the
spin current in the spacer. It follows that the values of the spin current incident on and leaving the switching 
magnet  
can both be determined from the profile of the spin current in the switching magnet. This is shown in Fig.7 for 
the situation when the polarizing magnet is a half-metal (the minority-spin band is empty) but the Fermi level 
in the switching 
magnet intersects both the majority and minority-spin bands.
\begin{figure}
\includegraphics[width=0.4\textwidth]{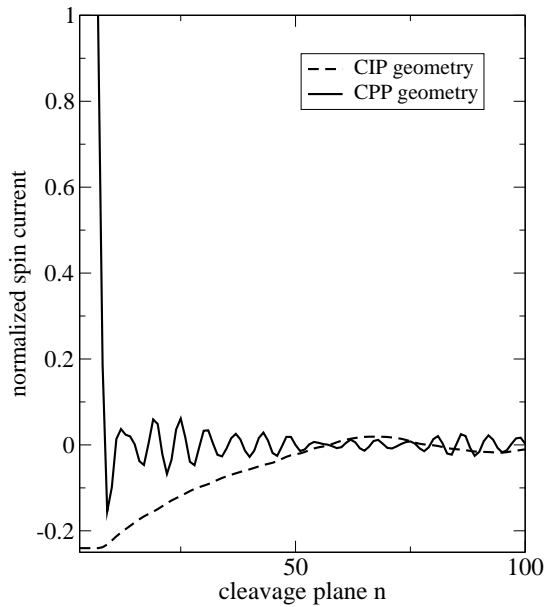}
\caption{\footnotesize Dependence of the spin current on the position $n$ of the cleavage plane in the
switching magnet for a CIP and a CPP system. The on-site potential parameters are $V^{\uparrow}=2.1$, $V^{\downarrow}=5.1$
for the PM, $V^{\uparrow}=V^{\downarrow}=2.1$ for the spacer and
$V^{\uparrow}=2.1$, $V^{\downarrow}=2.9$ for the SM.
The lead/magnet height is 10/10 atomic planes.} 
\end{figure}
There are two interesting features seen in Fig.7. First of all we note that in the CPP geometry only majority-spin
carriers can pass through a half-metallic polarizing magnet and, therefore, the spin polarization of the current 
incident on the switching magnet is 100\% and in the direction of the spin of the majority-spin carriers. 
On the other hand, 
the spin polarization in the CIP geometry is much smaller,
only about 25\%. The second interesting feature is that the spin polarization of the current in the CIP geometry 
has a sign opposite to that in the CPP geometry. This can be most easily understood in our special case of a
half-metallic polarizing magnet whose majority-spin band matches exactly the bands of either spin in the
nonmagnetic spacer. Minority-spin carriers, which cannot penetrate the polarizing magnet, travel as if in a
perfect slab without being scattered from the region in which the polarizing magnet is located. On the other hand,
majority-spin carriers which can easily penetrate the polarizing magnet region are strongly scattered by the
geometrical inhomogeneity of that region, which strongly reduces but does not suppress completely their current flow.
We thus do not expect the spin polarization to be complete. Moreover, the current of the 
minority-spin carriers is larger than that of the majority-spin carriers and the sign of the
spin current polarization is thus reversed.
%
\section{Results for Co/Cu lateral CPIP system.}
%
\noindent
Our model calculations for a single-orbital tight-binding band indicate that the absorption of the spin current
by a lateral magnet in the CPIP (CIP) geometry is as efficient as in the standard CPP geometry. To confirm 
that these
results remain valid for a fully realistic system, we have made  calculations of the spin
current profile in a cobalt switching magnet whose interface with a nonmagnetic copper spacer is parallel to
the current flow (CPIP geometry illustrated in Fig.2a). We used in these calculations a semiinfinite fcc Co 
sheet of height 4 and 8 atomic planes as a polarizing magnet. The switching magnet was a sheet of Co of 
height 4 (8) atomic planes deposited on a Cu lead whose height was also 4 (8) atomic planes. The crystal
orientation of the Co and Cu sheets was (001). Both Co and Cu sheets were described by a fully realistic
multiorbital tight-binding model with tight-binding parameters fitted to the results of first-principles band
structure calculations (see Ref.~\onlinecite{ref9}). The magnetization of the polarizing Co magnet was taken to be
in the x direction and that of the switching Co magnet was in the z direction. As in our
one-band model calculations, the Co/Cu CPIP system was grown in real space and the spin current was evaluated
without any approximations from the Keldysh formula (1). It should be noted that for a system with 8+8 atomic
sheets, all the matrices in Eq.(1) have size $(36\times 16)\times (36\times 16)$, which makes the evaluation of the spin current
computationally very demanding. Hence our restriction to the maximum size of 8+8 atomic sheets. The dependence
of the CPIP in-plane spin current on the position $n$ of the cleavage plane in the Co switching magnet is shown
in Fig.8. For comparison, the CPP spin current is also shown in Fig.8 (continuous line). In the CPP geometry,
\begin{figure}
\includegraphics[width=0.45\textwidth]{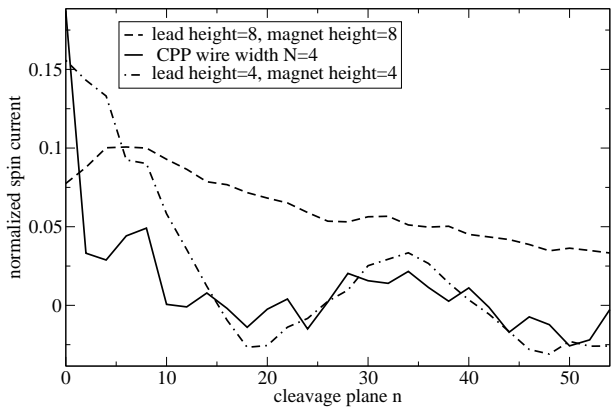}
\caption{\footnotesize Dependence of the spin current on the position $n$ of the cleavage plane in the
cobalt switching magnet.} 
\end{figure}
the Co polarizing magnet, the Cu spacer, and the Co switching magnet were all sheets of 4 atomic planes. 
It can be seen from Fig.8 that in the
case of the 4+4 CPIP system the absorption of the spin current is as fast as in the conventional CPP
geometry. The long-period oscillations of the spin current in the CPIP and CPP geometry are very similar but we
can see in the CPP geometry an additional short oscillation period which is not present in the CPIP geometry.
The absorption of the spin current for the 8+8 CPIP system is slower but, nevertheless, more than two thirds of
the spin current are absorbed over 50 atomic planes. Our results for realistic Co/Cu systems thus confirm the
viability of a setup with a lateral switching magnet, i.e. the CPIP geometry in which the current flows 
parallel to the spacer/switching magnet interface.
%
\section{Conclusions.}
%
\noindent
Using the nonequilibrium Keldysh theory, we have investigated theoretically two geometries for current 
induced switching of magnetization in
which the current flows parallel to the magnet/nonmagnet interface. In the first geometry the current 
is perpendicular to 
the polarizing magnet/spacer interface but parallel to the spacer/switching magnet interface (CPIP). 
In the second
geometry the current is parallel to both the polarizing magnet/spacer and spacer/switching magnet
interfaces (CIP). Our calculations for a single-orbital tight binding model indicate that the spin current  
flowing parallel to the switching magnet/spacer interface can be absorbed by a lateral switching magnet 
as efficiently as in the traditional CPP geometry. We have confirmed that the results of such model calculations 
in the CPIP geometry are also valid for
experimentally relevant Co/Cu CPIP system described by fully realistic tight binding bands fitted to 
an ab initio band structure. Our results show that almost complete absorption of the incident spin current 
by a lateral switching magnet (magnetic dot) occurs when the lateral dimensions of the switching magnet are
of the order of 50-100 interatomic distances, i.e., about 20nm. The numerical results are supported by an
analytical stationary phase argument which indicates that the decay of the spin current in a lateral switching magnet
should not be slower than $1/\sqrt{n}$, where $n$ is the lateral size of the magnet measured in the units of 
interatomic spacing. Hence about 90\% spin current absorption should be achieved by a magnet of a lateral size 
of about 20nm.
Moreover, to achieve full absorption of the spin current 
(maximum spin-transfer torque), the height of a lateral switching magnet can be as
small as a few atomic planes. It follows that the total volume of the switching magnet in the CPIP (CIP) geometry
can be even smaller than that in the traditional CPP geometry using magnetic nanopillars. This indicates that
current-induced switching and microwave generation in the CPIP geometry should be feasible. We have also demonstrated
that strong spin current absorption in the CPIP/CIP geometry is not spoilt by the presence of a rough interface
between the switching magnet and  nonmagnetic spacer. 

We find that the polarization achieved using a lateral magnet in the CIP
geometry is only about 25\% of that in the traditional CPP geometry. The CPIP geometry is thus preferable but 
CIP could be still usable with a stronger current.

Finally, we wish to make contact with the recent experiment, see Ref.~\onlinecite{ref55}, in which the so called
pure-spin-current-induced magnetization switching had been demonstrated. In the experimental setup of Ref.~\onlinecite{ref55}
the current was spin polarized by passing it through a magnet (current perpendicular to magnet/spacer interface) 
but the resultant spin current was absorbed by a lateral magnet (current parallel to magnet/spacer interface). 
The experimental setup of Ref.~\onlinecite{ref55} is thus topologically equivalent to our CPIP geometry. 
\begin{acknowledgments}
\noindent
We are grateful to the UK Engineering and Physical Sciences Research Council for financial support
within the framework of the Spin@RT Consortium and to the
members of the Consortium for stimulating discussions. 
\end{acknowledgments}


\begin{references}
\bibitem{ref1}
F.J. Albert, J.A. Katine, R.A. Buhrman, and D.C. Ralph, Appl. Phys. Lett. {\bf 77}, 3809 (2000).
\bibitem{ref2}
J.A. Katine, F.J. Albert, R.A. Buhrman, E.B. Meyers, and D.C. Ralph, Phys. Rev. Lett. {\bf 84}, 3149 (2000).
\bibitem{ref3}
S.I. Kiselev, J.C. Sankey, I.N. Krivorotov, N.C. Emley, R.J. Schoelkopf, R.A. Buhrman, and D.C. Ralph, 
Nature {425}, 380 (2003).
\bibitem{ref4}
S. Urazhdin, Norman O. Birge, W.P. Pratt, Jr., and J. Bass, Phys. Rev. Lett. {\bf 91}, 146803 (2003).
\bibitem{ref5}
M.R. Pufall, W.H. Rippard, Shehzaad. Kaka, S.E. Russek, T.J. Silva, Jordan. Katine, and Matt. Carey, Phys. Rev. B. 
{\bf 69}, 214409 (2004).
\bibitem{ref51}
D.M. Edwards and J. Mathon, J. Phys.: Condens. Matter {\bf 19}, 165210 (2007).
\bibitem{ref55}
Tao Yang, Takashi Kimura, and Yoshichika Otani, Nature Physics {\bf 4}, 851 (2008).
\bibitem{bauer}
A. Brataas, G.E.W. Bauer, and P.J. Kelly, Phys. Rep. {\bf 427}, 157 (2006).
\bibitem{ref6}
L.V. Keldysh, Soviet Physics JETP {\bf 20}, 1018 (1965).
\bibitem{ref7}
C. Caroli, R. Combescot, P. Nozieres, and D. Saint-James, J. Phys. C {\bf4}, 916 (1971).
\bibitem{ref8}
D.M. Edwards, F. Federici, J. Mathon, and A. Umerski, Phys. Rev. B {\bf 71}, 054407 (2005).
\bibitem{ref9}
J. Mathon, Murielle Villeret, A. Umerski, R.B. Muniz, J. d'Albuquerque e Castro, and D.M. Edwards, 
Phys. Rev. B {\bf 56}, 11797 (1997).
\bibitem{andrey}
A. Umerski, Phys. Rev. B {\bf 55}, 5266 (1997).
\bibitem{stiles}
M.D. Stiles and A. Zangwill, Phys. Rev. B {\bf 66}, 014407 (2002).
\bibitem{itoh}
J. Mathon, Murielle Villeret, and H. Itoh, Phys. Rev. B {\bf 52}, R6983 (1995).
\end{references}
\end{document}